# Tuning structure and rheology of silica-latex nanocomposites with the molecular weight of matrix chains: a coupled SAXS-TEM-simulation approach

Amélie Banc[1,2], Anne-Caroline Genix[1,2*], Mathieu Chirat[1,2], Christelle Dupas[1,2], Sylvain Caillol[3], Michael Sztucki[4], Julian Oberdisse[1,2]

[1] *Université Montpellier 2, Laboratoire Charles Coulomb UMR 5221, F-34 095, Montpellier, France*
[2] *CNRS, Laboratoire Charles Coulomb UMR 5221, F-34 095, Montpellier, France*
[3] *Institut Charles Gerhardt Montpellier, UMR 5253 CNRS-UM2-ENSCM-UM1, Equipe I.A.M., 8 rue de l'Ecole Normale, F-34 296 Montpellier Cedex 5, France*
[4] *European Synchrotron Radiation Facility, ESRF, 6 rue Jules Horowitz, BP 220, F-38 043, Grenoble, Cedex 9, France*

**Revised version**

**ABSTRACT**

The structure of silica-latex nanocomposites of three matrix chain masses (20, 50, and 160 kg/mol of poly(ethyl methacrylate)) are studied using a SAXS/TEM approach, coupled via Monte Carlo simulations of scattering of fully polydisperse silica nanoparticle aggregates. At low silica concentrations (1%v), the impact of the matrix chain mass on the structure is quantified in terms of the aggregation number distribution function, highest mass leading to individual dispersion, whereas the lower masses favor the formation of small aggregates. Both simulations for SAXS and TEM give compatible aggregate compacities around 10%v, indicating that the construction algorithm for aggregates is realistic. Our results on structure are rationalized in terms of the critical collision time between nanoparticles due to diffusion in viscous matrices. At higher concentrations, aggregates overlap and form a percolated network, with a smaller and lighter mesh in presence of high mass polymers. The linear rheology is investigated with oscillatory shear experiments. It shows a feature related to the silica structure at low frequencies, the amplitude of which can be described by two power laws separated by the percolation threshold of aggregates.

1. **INTRODUCTION**

Polymer nanocomposites are formed by incorporation of nanoparticles (NP) into polymer matrices, usually with the aim of improving material properties [1-4]. Among the latter, the mechanical or rheological behavior is found to depend strongly on microstructure, i.e. on the spatial arrangement of the NPs in the matrix [3, 5-8]. Nanocomposites, in particular model systems obtained by mixing polymer and NPs in solution followed by solvent casting have been abundantly studied in the literature [9-12]. These systems allow for a large variation of experimental parameters, like matrix polymer chain mass [13, 14], or grafting chemistry on NPs [12, 15]. At high enough temperatures, typically above the matrix $T_g$, the structure of systems can evolve if the NP mobility is sufficiently high. This effect has been exploited by several authors [9, 16-18]. Jia et al [17], e.g., have studied NP aggregation as a function of annealing temperature and matrix molecular mass, finding higher aggregation at elevated temperatures and low masses. This effect is related to the matrix viscosity, which controls the NP diffusivity. In most of the references cited above, structural analysis of complex assemblies or dispersions is performed by small angle scattering or electron microscopy techniques. The latter are rarely coupled [19], and only a few quantitative comparisons of reciprocal and direct space methods exist, e.g. [5, 20].

The latex-route is another way to obtain polymer nanocomposites with controllable structures. Latex film formation is well-understood [21, 22], and the dissolution of latex beads in presence or absence of silica NPs has been recently followed in detail [23]. The dispersion of the NPs in the latex matrix is primarily a consequence of the interactions between the beads while still in aqueous suspension. Triggering repulsions by pH has been used by us before to change aggregation in the final nanocomposite [24, 25], and thus reinforcement [26]. As with solvent-cast nanocomposites, reorganization on accessible time scales of the NPs after evaporation of the aqueous solvent is possible if the mobility of the NPs in the matrix is high enough, the latter being controlled by matrix viscosity, and thus matrix chain mass [27]. To our best knowledge, there is no study on the effect of chain mass on filler reorganization in latex films, which is the purpose of the present contribution.

The outline of this article is the following. First an analysis of transmission electron microscopy (TEM) pictures in terms of distribution functions of aggregation numbers is presented, for nanocomposites made with matrix chains of three different masses. Then, a small angle X-ray scattering (SAXS) analysis is quantitatively coupled to the TEM results by introducing the aggregate distributions in a fully polydisperse Monte Carlo simulation of scattering of aggregates. Apart from a recent study done on a different system [19], we are not aware of such quantitative combined studies. In the second part of the results section, we explore the rheological properties of the nanocomposites, and isolate the silica contribution before and after the percolation threshold. In the discussion, finally, the impact of the rheology on the structure, and of the structure on the rheology, is compared.

## 2. MATERIALS AND METHODS

**Silica nanoparticles:** The aqueous solution of silica NPs, Ludox TM40, was purchased from Aldrich and was characterized by SAXS after dilution in deionized water down to a volume fraction of 0.004. The scattered intensity was well fitted with a log-normal distribution of spherical objects ($R_{si}$ =14 nm and $\sigma$ = 11%, cf. SI). From this distribution, the average silica NP volume and the average projected surface area can be calculated: $V_{si}$ = 1.21 $10^4$ nm$^3$, and $A_{si}$ = 630 nm$^2$. Image analysis of TEM results gave very similar NP characteristics (cf. SI): $R_{si}$ = 13.1 nm and $\sigma$ = 12%.

**Poly(ethyl methacrylate) (PEMA) nanolatices:** Polymer NPs were synthesized by free radical emulsion polymerization using a semi-continuous batch method. The synthesis was adapted from similar protocols [28]. All materials were used as received. Ethyl methacrylate (EMA, Aldrich, 99% purity), ammonium persulfate (($NH_4$)$_2$$S_2$$O_8$), tert-dodecylmercaptan (TDM Aldrich, 98.5% purity) and sodium dihydrogenphosphate ($NaH_2PO_4$ - $H_2O$) were used as initiator, transfer agent and buffer, respectively. The surfactant used for stabilization was sodium dodecyl sulfate (SDS, >98.5% purity). Deionized water was used throughout. The semi-continuous emulsion polymerization was carried out at 80°C in a 250 ml glass reactor fitted with a reflux condenser, stainless-steel stirrer, temperature probe, argon inlet and two feed inlet tubes. The initial charge in the reactor was the surfactant, buffer, initiator and water. To remove oxygen, argon gas was bubbled through the solution into the reactor for 15 min under 250 rpm stirring. Once the temperature of the charge stabilized at 80°C, 0.5 mL of initiator solution (0.324 g in 2 ml of water) was injected into the flask and the feed streams were initiated. The first feed stream was a solution of initiator, surfactant and buffer in water (feeding rate = 0.39 g/min), and the second was the monomer-transfer agent mixture (feeding

rate = 0.194 g/min to maintain starved conditions). The monomer and aqueous feeds were calculated to finish the addition in 1 h. Subsequently, the polymerization was continued in batch for 1 h. Different chain masses (from 20 to 160 kg/mol) were generated by varying the amount of transfer agent. In our nomenclature PEMA_M corresponds to a sample with a M kg/mol matrix. PEMA160 was synthetized without transfer agent, whereas PEMA50 and PEMA20 were obtained using 0.0059 and 0.0108 TDM/EMA mass ratios, respectively. The polymer mass content of nanolatex was 10%w.

Nanolatex particle sizes were characterized by dynamic light scattering using a cumulant analysis. Polymeric particle diameters were comprised between 20 and 30 nm. After drying and solubilization in tetrahydrofuran, molecular weight characterization of polymeric chains was done by gel permeation chromatography (GPC) using PMMA standards for calibration. Polydispersity indexes were all less than 1.9. Although it is possible to use controlled radical polymerization to considerably decrease polydispersity indices [29, 30], the wide range of masses studied here - from a few entanglement masses of PEMA (between 6 and 8 kg/mol according to linear rheology, compatible with ref. [31]) to almost twenty times more - made such a refined approach unnecessary. Based on carbonyl peaks analysis of $^{13}$C NMR spectra obtained in CDCl$_3$, the tacticity of PEMA chains was measured: 77.5% syndiotactic and 22.5% isotactic. Characteristics of initial colloidal solutions are summarized in Table 1.

|  | **PEMA20** | **PEMA50** | **PEMA160** | **Silica TM40** |
|---|---|---|---|---|
| Nanoparticle diameter $D_p$ (nm) | 20 | 26 | 19 | 28 |
| $D_p$ polydispersity index | 0.3 | 0.3 | 0.3 | 0.11 |
| Molecular weight $M_w$ (kg/mol) | 18.5 | 51.6 | 159 | - |
| Polydispersity index $M_w/M_n$ | 1.7 | 1.6 | 1.9 | - |
| $T_g$ (°C) | 65 | 74 | 76 | - |

**Table 1:** Latex and silica nanoparticle characteristics.

Note that there are some minor differences in the nanolatex sizes. As this article focuses on the effect of polymer molecular weight, and a strong difference between PEMA160 and PEMA20 (which have virtually the same diameters) is found, and no difference between PEMA20 and PEMA50 (which differ slightly in diameter), it can be safely concluded that this parameter is not relevant for the present study. The glass-transition temperature of PEMA matrices has been determined by Differential Scanning Calorimetry (DSC) at 20 K/min and increases from 65°C to 76°C with increasing molecular weight following the Flory-Fox equation [32] (see Table 1 and SI for details).

**Nanocomposite film preparation:** Prior to preparation of nanocomposite films by solvent casting in a Teflon mold, colloidal solutions were extensively dialyzed against deionized water to remove residual reactants and salts until obtaining a conductivity inferior to 10 µS/cm. Mass contents of aqueous solutions were determined by the gravimetric method. After dialysis, the nanolatex mass contents were about 5% w/w and silica colloidal solution was 15% w/w. Considering measured densities of silica and PEMA ($\rho_{si}$ = 2.25 g/cm$^3$ and $\rho_{PEMA}$ = 1.13 g/cm$^3$, respectively), mixtures of solutions were prepared. Mixtures were degased overnight in a dessicator, whereas the Teflon mold was degased into the oven at 80°C under vacuum. Mixtures were dried about 8 hours in an oven at 80°C to obtain transparent films. Subsequently, to ensure latex bead dissolution [23], samples were annealed for seven days at 180°C under vacuum. During this annealing, residuals surfactants degraded and provided a slightly brown coloring to samples. GPC analysis showed that polymer molecular weights were not affected by this procedure (see SI). Note that similar to many acrylates, PEMA has favorable interactions by hydrogen bonding with bare silica[33, 34], which should be independent of polymer mass.

**Nanocomposite film characterization:** Thermogravimetric analysis (TGA) was performed to check silica volume fraction $\Phi_{si}$ of nanocomposite samples. 20 mg samples were heated from 30°C to 650°C at a rate of 30 K/min under nitrogen. From the plot of sample weight vs. temperature, the sample weight fraction W remaining after thermal decomposition was determined. For pure PEMA, this is denoted $W_{PEMA}$ (resp. $W_{NC}$ for nanocomposites). To correct for the small amount of incombustibles present in the pure PEMA ($W_{PEMA}$ of the order of 2%), the silica fraction in nanocomposites was determined by:

$$\Phi_{si} = \frac{W_{NC} - W_{PEMA}(1-\Phi_N)}{(W_{NC}-W_{PEMA}(1-\Phi_N))/\rho_{si} + (1-W_{NC}-W_{PEMA}(1-\Phi_N)))/\rho_{PEMA}} \quad (1)$$

The $\Phi_{si}$ values determined by TGA were in good agreement with the nominal volume fraction $\Phi_N$ as deduced from the polymer and silica weight used in the preparation. Only the nominal silica volume fractions will be given throughout this article ($\Phi_{si} \approx \Phi_N$). The glass-transition temperature of the PEMA composites did not show any significant variation with the silica fraction, in agreement with Moll et al [34] in this concentration range.

**Rheology:** The rheological response in the linear regime of nanocomposites was obtained with a stress-controlled rheometer AR2000, used in the strain-controlled mode in plate–plate geometry (20 mm diameter). Isothermal frequency sweeps at fixed low deformation level ($\gamma$ = 0.2%) were performed in the temperature range from 100°C to 180°C. Using the principle of time–temperature superposition, master curves of the storage modulus, $G'(\omega)$, and the loss modulus, $G''(\omega)$, were constructed at a reference temperature of 180°C based on frequency sweeps between $\omega$ = 0.01 and 600 rad/s. In order to evaluate the statistical relevance of the results, five identical samples for both the matrix and 10% nanocomposites were formulated independently and characterized in rheology. The error bars for the moduli are about 18% with 95% confidence intervals, which is comparable to the symbol size and thus rather low with respect to the many orders of magnitude covered by G' and G''. The horizontal shift factors, $a_T$, obtained from the master curve construction were not found to evolve from the matrix to the highest silica fraction for the three masses. Their evolution with the inverse of temperature can be well described with the classical Williams–Landel–Ferry (WLF) equation[35] using the PEMA coefficients from ref. [36] ($C_1$ = 6.3 and $C_2$ = 184 K at our reference temperature, see SI). It means that the temperature dependence of the relaxation process probed here is not significantly modified by the introduction of filler, in spite of strong variation of the G' and G'' shapes. Such behavior was already observed in the literature for nanocomposite systems.[5]

**Structural analysis of nanocomposites:** Samples for TEM were prepared by immersing a thin strip of sample in an epoxy resin (EPON 812) and curing it at 60°C for 72h. After resin polymerization, sections with a nominal thickness of 70 nm were cut with an ultramicrotome (Leica Ultracut) and placed on TEM grids (Formvar carbon-coated Cu grids, EMS). Slices were observed with a 1200EX2 Jeol TEM at 100 kV. Images were captured with a Quemesa SIS Olympus numerical camera equipped with an 11 MPixels CCD detector. Representative images as those shown in this article were obtained with a 20 000 magnification. For the 1% nanocomposites, image analysis was performed on 3000 magnification pictures in order to achieve better statistics via observation of a higher number of aggregates. The grain analysis tool of Gwyddion Software was used to identify the silica NPs by intensity thresholding, and to determine the distribution of polymer-free areas occupied by aggregates. The resolution of the images is 200 dpi (magn. 20000, pixel size 0.72 x 0.72 nm$^2$) for the TEM data shown in this article. Images used for quantitative analysis of many aggregates have a pixel size of 4.5 x 4.5 nm$^2$). An average NP covers ca. 27 pixels, or about 6 pixels in diameter, which gives us an errorbar estimate of ±1 pixel, or a maximum of ± ca. 30% of a single NP area. Due to the measurement of many NPs, the error in aggregation number associated with the precision of the TEM pictures is probably a factor of ten lower, i.e. ca 3%. Small angle X-ray scattering (SAXS) was performed on beamline ID2 at the European Synchrotron Radiation Facility (ESRF, Grenoble FR) at a wavelength of 1 Å with a sample to detector distance of 2.5 m, yielding a total q-range from 0.001 to 0.15 Å$^{-1}$. The scattering cross section per unit sample volume $d\Sigma/d\Omega$ (in cm$^{-1}$) – which we term scattered intensity I(q) – was obtained by using standard procedures including background subtraction and calibration given by ESRF.

**Monte Carlo simulations of intra-aggregate structure factors:** The scattered intensity of an isotropically averaged single aggregate containing $N_{agg}$ *identical* nanoparticles is given by the product of the form factor P(q) and the intra-aggregate structure factor $S_{intra}(q)$.[37] The former represents the scattering from individual spheres, whereas the latter expresses the Fourier transform of the positional correlations of the NPs in an aggregate. The isotropically averaged contribution to the partial structure factor due to two spherically symmetric beads can be calculated using the Debye formula [38]

$$S_{ij}(q) = \frac{\sin(q(r_j-r_i))}{q(r_j-r_i)} \qquad (2a)$$

where $r_i$ and $r_j$ are the bead positions (center of mass). Summing all contributions gives

$$S_{intra}(q) = 1 + \frac{2}{N_{agg}} \sum_{i>j}^{N_{agg}} S_{ij}(q) \qquad (2b)$$

where $N_{agg}$ is the number of beads per aggregate. Obvious limits of $S_{intra}(q)$ are $N_{agg}$ at low q, and one at high q. The detailed positions of the NPs need to be fixed by some aggregate construction algorithm. Here, we simply aggregate the $N_{agg}$ nanoparticles randomly starting with a seed particle and sticking one NP after the other to the growing aggregate, on a random position, provided that there is no collision. Note that additional NPs are thus added on any point at the surface of any previously positioned NP of the aggregate, thereby possibly creating branches. Moreover, the only input for the structure factor in eq. (2a) being the distance, it is not even necessary for NPs to be topologically connected, as close interparticle distances may also occur without sticking in the random aggregation process. In the absence of any a priori knowledge on the mechanisms of aggregate construction, only connectivity (via the minimum distance between NPs, defined by their excluded volume) and $N_{agg}$ are thus imposed. Two features of $S_{intra}$ are thereby constrained: the low-q value of $S_{intra}$ describing the total mass of the aggregate, and the position of the NP-NP correlation peak.

Adding *polydispersity* in size for the nanoparticles requires to weight properly each partial correlation by the proper mass, i.e. using $P_i(q)$ and $P_j(q)$, and the use of a simple product ($P \times S_{intra}$) is no more valid [39]. In the case of assemblies of N independent aggregates (i.e., measured at high dilution) of different $N_{agg}$, the total scattered intensity is given by a weighted average and reads [19]

$$I(q) = \frac{\Phi_{si} \Delta\rho^2}{V_{si\,total}} \sum_N \sum_{i,j=1}^{N_{agg}} V_i V_j \sqrt{P_i(q) P_j(q)} S_{ij}(q) \qquad (3)$$

where $V_i$ and $V_j$ are the corresponding silica volumes of the NPs i and j. In eq. (3), the first sum runs over the total number of aggregates N. There, we used the distribution function in $N_{agg}$ obtained from the electron microscopy. The averaging sum accounts correctly for the contributions of aggregates of different mass, with a stronger weight for bigger aggregates. One may note that using the monodisperse version, eq. (2b), has been tried out initially. This amounts to taking into account the distribution of center-to-center distances introduced by polydispersity, but not the weighting. It was shown that for our low polydispersity samples ($\sigma = 11\%$) this estimation was acceptable for the low-q scattering, but induced clearly visible artefacts in the region of the form factor oscillations, i.e. at rather low intensities.

## 3. RESULTS

### 3.1 Nanocomposite structure

The structure of the silica NPs in the PEMA-nanocomposites has been studied as a function of silica volume fraction ($\Phi_{si}$ = 1, 3, 10%v) and matrix chain mass (20, 50, 160 kg/mol, see Table 1) by TEM. Representative results are shown in Figure 1. Quantitative analysis as described below has been performed on several pictures at six times lower magnification. From the different apparent silica densities, some difference in the real thickness of the slices can be deduced. For 1%v samples, the thickness is 70±20 nm. At 10%v, the thickness decreases from about 110±10 nm for PEMA20 to ca. 65±5 nm for PEMA50 and PEMA160.

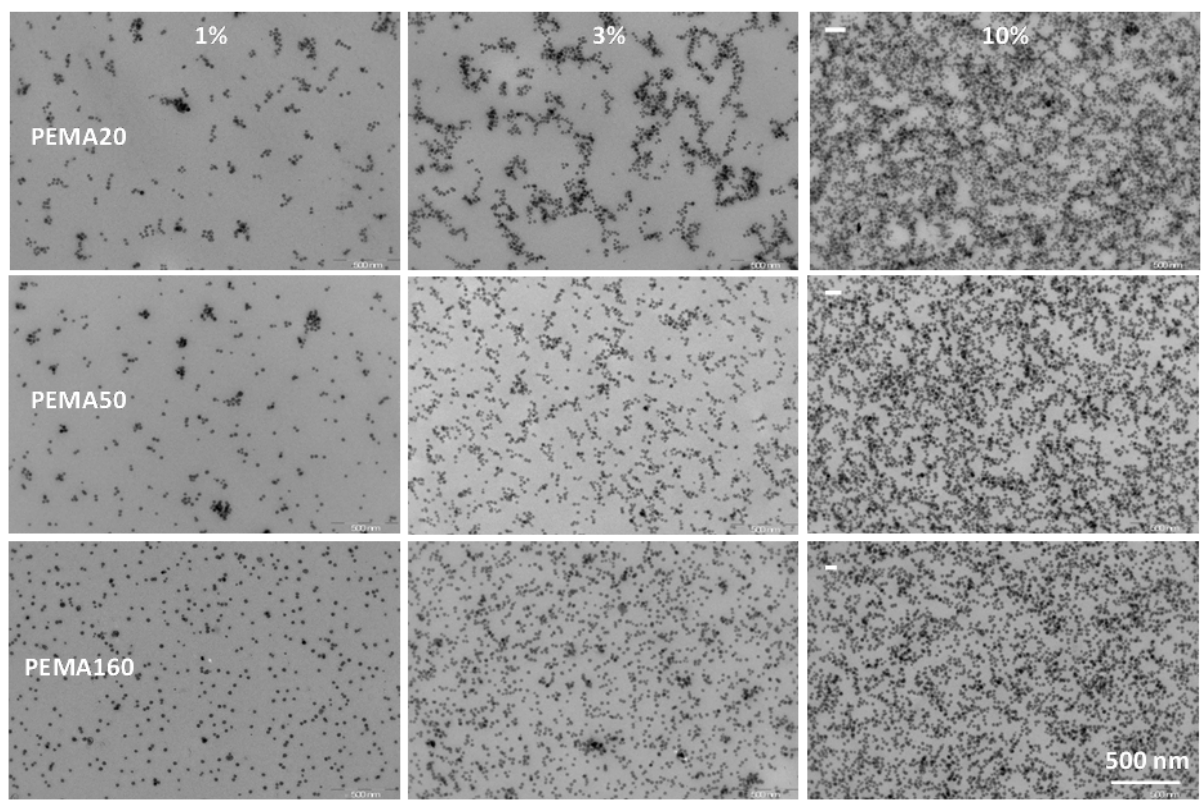

**Figure 1:** TEM images of nanocomposites made with PEMA20 (upper line), PEMA50 (intermediate line) and PEMA160 (lower line), at three silica volume fractions: 1%v (left), 3%v (center) and 10%v (right). In spite of a fixed nominal thickness of the slices (70 nm), some variations are observed as discussed in the text. White horizontal bars in the right column of images indicate the characteristic size measured for 10% nanocomposites by SAXS.

The influence of silica volume fraction is readily seen by comparing the three columns in Figure 1. For 1%v, shown in the left column, the vast majority of NPs are well-dispersed, i.e. individually or in small aggregates at all polymer masses. Judging from the absence of any aggregate in the lowest image (160 kg/mol), the dispersion is best at the highest mass. At 3%v, in the series shown in the column in the middle, the beginning of network formation is observed. At 20 kg/mol, thick and lengthy aggregates are visible, but these branches do not form a percolating network yet. Due to the heterogeneous distribution, it is difficult to estimate the lateral branch size, but a typical number of two to four NPs, corresponding to 60 to 120 nm, can be put forward. At 50 kg/mol, and similarly at 160 kg/mol, much less aggregation is evidenced: branches become thinner, of typical lateral dimension one or two NPs, and break-up into shorter aggregates. Thus, at 3%v, the higher mass moves the system away from percolation. At the highest silica volume fraction of 10%v, finally, in the TEM pictures shown in the right-hand column in Figure 1, a thick network is found at the low polymer mass of 20 kg/mol. The typical silica network mesh size is difficult to estimate from the TEM pictures, but from the typical width of the silica-free holes or channels it is seen to be around 150 nm. This value can be compared to the mesh size extracted from the SAXS measurements below, and the corresponding size bar is superimposed in Figure 1. A thinner network is observed at 50 kg/mol, with the same typical holes and thus approximate silica mesh size. At the highest mass of 160 kg/mol, the presence of still many individual NPs together with polymer channels indicates that a silica network is under formation, of again roughly the same typical mesh size as with lower masses. Summarizing the visual inspection of the TEM-pictures shown in Figure 1, (i) an increase in silica volume fraction favors the formation of a network. (ii) For a given silica concentration, samples appear to be better dispersed in nanocomposites made with higher polymer masses.

Using image analysis, it is possible to quantify the extent of aggregation, at least for the lowest volume fraction, where individual aggregates and NPs are easily recognized. Two procedures were used to estimate minimum and maximum values for the number of NPs per aggregate, respectively $N_{agg}^{min}$ and $N_{agg}^{max}$. The hypothesis for the $N_{agg}^{min}$ calculation is that the number of superimposing projections of NPs in the plane of observation is negligible at low volume fractions and the whole silica aggregate is contained in the TEM sample slice of thickness e. Hence $N_{agg}^{min}$ can be calculated dividing the area occupied by silica inside aggregates $A_{agg}$ by the average area occupied by one silica nanoparticle $A_{si}$ as determined from the NP log-normal size distribution given in the materials section:

$$N_{agg}^{min} = \frac{A_{agg}}{A_{si}} \tag{4}$$

Note that an estimate of the aggregation number based on numerical simulations of fractal aggregates has been proposed [40] and applied in the literature [5]. The local silica volume fraction inside aggregates, which defines the compacity $\kappa$, can be estimated for each aggregate from TEM images by dividing the observed silica volume by the aggregate volume in the slice. Each such aggregate volume was calculated by multiplying the sample slice thickness e with the observed surface area $\pi R_{agg}^2$, where the radius $R_{agg}$ is the one of the smallest disc in which the aggregate can be contained.

$$\kappa^{TEM} = \frac{N_{agg}^{min} V_{si}}{e \pi R_{agg}^2} \tag{5}$$

$V_{si}$ denotes the average NP volume. For very small aggregation numbers $N_{agg}$ = 1 and 2, the compacity was set to its geometrical value of 100% and 25%, respectively. For $N_{agg}^{min} \geq 3$, we have determined the average over $\kappa^{TEM}$ for hundreds of aggregates in pictures with typically 10 000 NPs, giving on average $\kappa^{TEM} \approx 10\%$ for the three polymer masses. This value as well as its evolution with aggregate size will be compared to the simulation results later in this article.

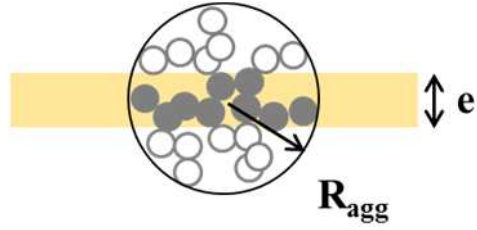

**Figure 2:** Schematic representation of silica NP aggregate in TEM slice. Solid NPs in the film thickness e are observed by TEM and define $R_{agg}$ and $N_{agg}^{min}$. Empty NPs have been estimated to be part of the initial aggregate but removed through the cutting procedure. They contribute to the estimation of $N_{agg}^{max}$ as explained in the text.

Aggregates may have been cut during slicing, and we can use this compacity $\kappa^{TEM}$ and the observed 2D-size $R_{agg}$ to estimate the aggregation number before slicing, $N_{agg}^{max}$:

$$N_{agg}^{max} = \frac{\frac{4}{3}\pi R_{agg}^3 \kappa^{TEM}}{V_{si}} \qquad (6)$$

Here, we assume that the observed 2D-size is representative of the initial 3D-size of aggregates before slicing, as illustrated in Figure 2. The limitation due to the two-dimensional projection of three-dimensional aggregates is obvious: there is no reason for aggregates to be of spherical symmetry, and therefore the diameter $2R_{agg}$ as determined from the projection is only an estimation lying somewhere between the minimal and the maximal extension of the aggregates. The aggregates were automatically identified in low magnification pictures (×3000) displaying more than 5000 objects, using the intensity threshold tool of Gwyddion software. $A_{agg}$ and $R_{agg}$ were calculated using the grain analysis tool of the software.

Note that the aggregation numbers defined by eqs. (4) and (6) are approximations of the real ones, due to the use of the average particle area instead of the individual ones. For the same reason, fractional values of aggregation numbers can be obtained. Therefore, a binning operation into integer values of the aggregation number was performed on the $N_{agg}$-distributions. This was necessary for the quantitative cross-check between direct-space (TEM) and reciprocal space (SAXS) which we will develop below, using the integer distribution as input for the Monte Carlo simulations. The results for the distribution $f_{si}$ of $N_{agg}^{min}$ and $N_{agg}^{max}$ are shown in Figure 3, for the lowest silica volume fraction (1%v), and the three different polymer masses. $f_{si}(N_{agg})$ is the fraction of NPs present in the form of aggregates of aggregation number $N_{agg}$.

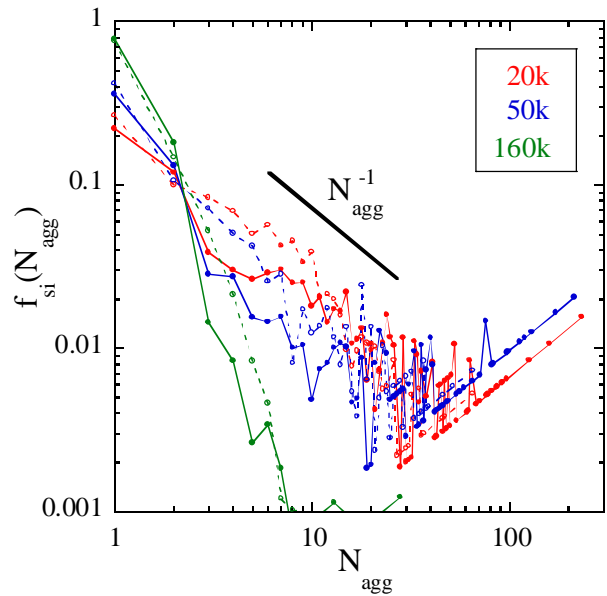

**Figure 3:** Distribution of $N_{agg}^{min}$ (empty symbols) and $N_{agg}^{max}$ (full symbols) obtained from the image analysis of 1%v nanocomposites of different matrix molecular weight.

The distributions of the aggregation numbers shown in Figure 3 show that the overwhelming majority of the NPs is present individually or in the form of very small aggregates. At 160 kg/mol, e.g., about 80% of all NPs are single beads. This fraction decreases to between 20 and 40% for lower masses. A tail up to large $N_{agg}$ is evidenced for the lowest chain masses, which can be roughly described by a $N_{agg}^{-1}$ power law. The decrease is much stronger for PEMA160, where the power law exponent approaches -3. Such a steep power law leads to the virtual absence of any aggregates having more than ten beads in this TEM analysis. As a last comment, one may note the linear increase of the distribution at high $N_{agg}$. It corresponds to big aggregates existing only once, and their contribution increases with their mass. Given the low fraction of NPs involved in these aggregates, this increase is an artefact caused by the limited sample size, 2% of ca. 15000 NPs counted being the order of the highest $N_{agg}$.

We have combined the direct-space analysis by TEM with SAXS experiments. In reciprocal space, scattering experiments give a signal corresponding to the structure averaged over large sample areas. The latter are given by the X-ray beam size (typically 100 μm), as compared to the small piece of matter focused on in an electron microscope. Also, in SAXS, complete and unperturbed three-dimensional aggregates are observed. One of the drawbacks of SAXS is that data analysis is much more involved, but precious information can be extracted as we will see now. The reduced intensities $I(q)/\Phi_{si}$ of the series with PEMA20, PEMA50, and PEMA160, are plotted in Figure 4, with clear features. First, the high-q data show the same form factor oscillations as expected for identical silica NPs in all samples, minor deviations being due to matrix contributions as discussed below. Secondly, a different organization is measured in the low-q range in Figures 4a to 4c. There the reduced low-q intensity is seen to be highest for the smallest silica volume fraction (1%v), which we interpret as the form factor scattering of independent aggregates at this high dilution. A Monte Carlo analysis of this average form factor based on the $N_{agg}$-distribution of the TEM pictures will be proposed below for the 1%v series. As the volume fraction is increased, the interactions between aggregates induce a depression of the low-q intensity, indicating structuring of the aggregates in space due to repulsive interactions, and the emergence of a peak at intermediate wave vectors.

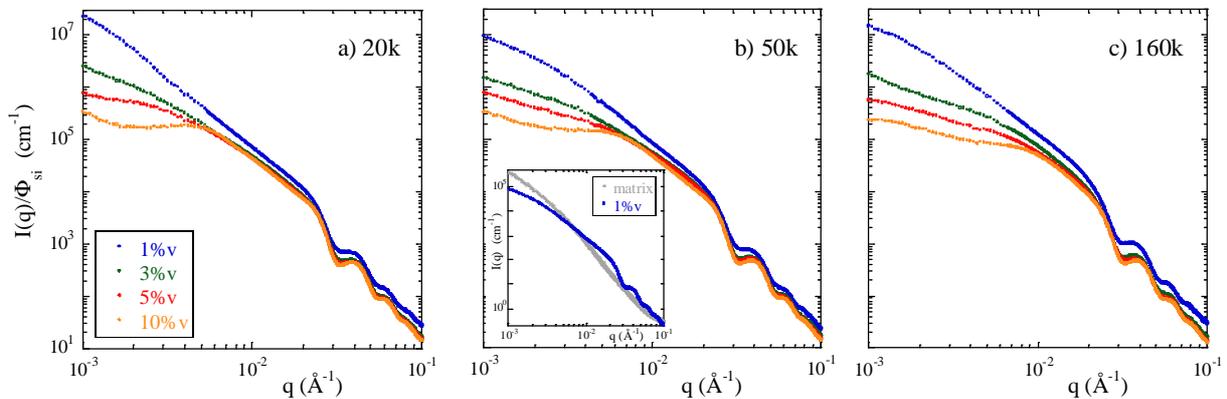

**Figure 4:** Reduced SAXS scattered intensity $I(q)/\Phi_{si}$ as a function of wave vector q of (a) PEMA20, (b) PEMA50, and (c) PEMA160 nanocomposites before matrix subtraction. In the inset: comparison of absolute intensities of a nanocomposite (50 kg/mol, 1%v) and its matrix.

A technical point concerns the subtraction of the pure matrix background. One sees in any of the graphs in Figure 4 that the matrix has the most important relative influence at the lowest $\Phi_{si}$ (not overlapping in the high-q range). An example is shown in the inset of Figure 4b. There the intensity $I(q)$ of PEMA50-1% nanocomposites is compared to its pure matrix intensity. The latter is found to be a monotonously decreasing function with q, reaching high enough intensities at low q to cross the 1%v-nanocomposite intensity around 3 10$^{-3}$ Å$^{-1}$. This illustrates the fact that the matrix properties change upon addition of silica, possibly modifying large-scale heterogeneities, and force us to introduce a low-q cut-off for a quantitative analysis of the data. Our procedure is to subtract a weighted matrix intensity such that the form factor oscillations of the NPs after subtraction – which correspond to the shape of the inorganic beads and are thus unaffected by any manipulation – are identical to the independently measured pure form factor given in the SI. We then ignore any intensity values below the cut-off. All data from Figure 5 on have undergone this procedure.

In Figure 5a, the same scattering data as in Figures 4a-c is regrouped for 10%v of silica in a single plot, after background subtraction. It allows following the characteristic distance of the silica network as given by the peak position q*. Our interpretation of the scattering data is suggested by the TEM-pictures, where network-like structures are found. First of all, there is an identical power law signature at intermediate angles, with $D_f = 2.5 \pm 0.05$, which speaks in favor of aggregation. As the aggregates are concentrated by increasing the silica volume fraction, they are pushed together, and at some stage the typical inter-aggregate distance becomes smaller than the size of the aggregates, and interpenetrated aggregates form the network visible in the TEM pictures.

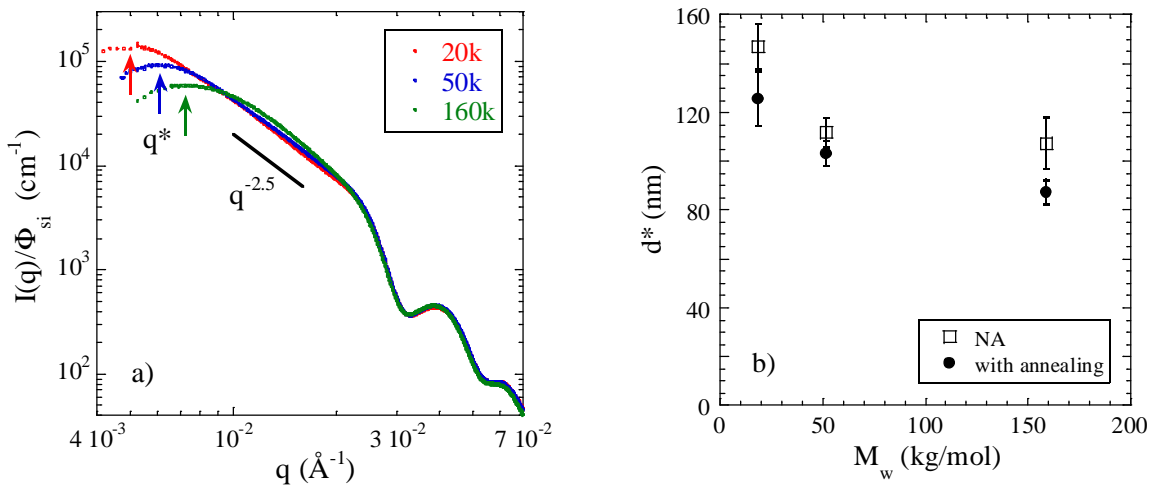

**Figure 5:** (a) Reduced SAXS-intensities of 10%v-nanocomposites with 20, 50 and 160 kg/mol matrices. Arrows indicate the peak position q*. (b) Silica mesh size $d^* = 2\pi/q^*$ for different chain masses, before (NA) and after annealing.

In the scattering in Figure 5a, the peak position is found to shift to higher q-vectors with higher chain masses, indicating the formation of a network of smaller characteristic distance given by d*=2π/q*. We identify this distance with the silica mesh size, and it is reported for the three chain masses in Figure 5b: d* is found to decrease with increasing chain mass, in agreement with the general evolution observed in Figure 1 towards a better dispersion at higher mass. Also in Figure 5b, the silica mesh size before annealing (i.e., directly after film formation) is reported for further discussion. Finally, for comparison with TEM, we have superimposed bars of length d* in Figure 1 as determined by SAXS. At 10%v, the higher the mass, the smaller and lighter the silica meshes.

In the discussion of Figure 4, we have underlined the fact that form factor scattering describing aggregates is observed at low silica concentrations. The difficulty is that the observed intensity is an average over many different independent aggregates, and it is impossible without additional knowledge to deconvolute I(q) into its unknown contributions from aggregates of different mass and shape from the scattering alone. However, we have quantitative estimations of the aggregation number distribution deduced from the TEM observations as shown in Figure 3. We have therefore set up a Monte Carlo simulation scheme to account for the scattering of aggregates obeying the experimental $N_{agg}$ distribution. In Figure 6, we compare its predictions to the observed intensities for 1%v-samples (PEMA20, 50 and 160, same data as Figure 4, now with background subtraction), i.e. only for the highest dilution where structure factors due to correlations between aggregates can be neglected. The experimental intensities after background subtraction follow a power law with $D_f$ = 1.7 ± 0.1 for the lower masses, indicative of aggregation and compatible with, e.g., diffusion limited cluster-cluster aggregation [41]. Comparison with the power law at 10%v given in Figure 5a indicates that the apparent fractal dimension increases with silica concentration, as already found by simulations in ref. [42, 43]. At the highest mass, the scattered intensity shows a different, form factor-like behavior, which suggests good dispersion, in agreement with the TEM picture.

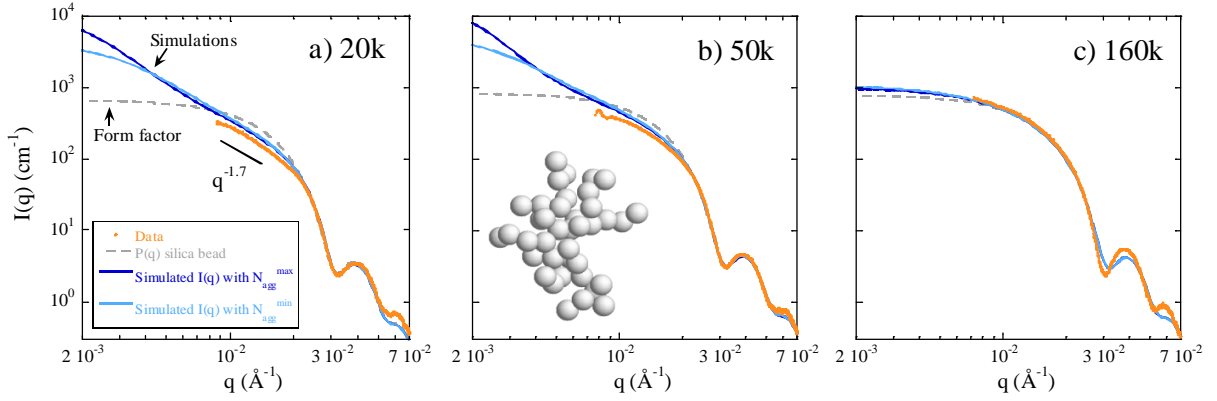

**Figure 6:** Comparison of experimental scattered intensities I(q) of 1%v-nanocomposites to simulations based on the different experimental $N_{agg}$ distributions as indicated in the legend. (a) PEMA20, (b) PEMA50, (c) PEMA160. The silica form factor is also plotted (dotted line), as well as a snapshot of a simulated aggregate ($N_{agg}$ = 40).

Due to our choice of the construction algorithm of aggregates as explained in the methods section, the simulated intensity in eq.(3) does not have any free parameter. For each polymer mass, the simulation has been run with the upper and lower estimations for $N_{agg}$, i.e. $N_{agg}^{max}$ and $N_{agg}^{min}$, respectively. A first striking result of Figure 6 is that both predictions do not differ significantly in the relevant q-range. Only at the lowest q-values, which are below the experimental cut-off, a discrepancy is observed. It is thus concluded that the range of aggregation numbers is correctly given by $N_{agg}^{min}$ and $N_{agg}^{max}$, thereby validating our quantitative analysis of the TEM pictures. A technical comment may be suited here. Aggregation is generally related to an increase of the low-q intensity, accompanied by a *decrease* with respect to the form factor of the intermediate-q intensity. Both features are visible in Figure 6. This decrease, known as the correlation hole, is a direct consequence of the excluded-volume correlations of neighboring and thus aggregated beads. Due to the cut-off at low-q, aggregation is discussed here in terms of an intensity decrease.

Next, the agreement between the experimental intensities and the calculated ones is remarkable. For the lowest chain mass, the low-q increase is correctly reproduced in Figure 6a, with only a slight mismatch below 0.02 Å$^{-1}$. This deviation indicates that the NP-NP correlation hole is more pronounced in real samples, which is probably due to a denser assembly on this scale. The nanocomposite made with PEMA50 has a very similar scattering signature, and is thus equally well fitted with the also quite similar $N_{agg}$-distribution. For the highest mass, the scattered intensity is quite different. It follows the pure bead form factor at low q, and this is again quantitatively reproduced by the model in Figure 6c. The overall agreement between the calculated curves and the experimental intensities for the three masses is due to three features. At the highest q, the form factor dominates the scattering. At intermediate q, the local density and thus coordination number seems to be correctly described by the aggregate construction proposed here, and there is only little room for improvement. At the lowest q, finally, the increase in intensity is related to the weighted average of the mass of aggregates, as one can deduce from the low-q limit of the intra-aggregate structure factor which is $N_{agg}$ for monodisperse distributions. To summarize, the quantitative agreement observed in Figures 6a to 6c shows that the $N_{agg}$-distributions extracted from the TEM pictures are trustworthy, and thus that the 160 kg/mol samples are significantly less aggregated than the shorter chain mass nanocomposites.

The difference in silica structure between low (20 and 50 kg/mol) and high (160 kg/mol) mass nanocomposites cannot be due to the drying stage during which the latex beads keep their colloidal character, irrespective of their internal structure and thus chain mass. It is therefore conjectured that the differences are due to the different kinetics once the water is evaporated, and we will come back to this point in the discussion. It is instructive to compare the structures directly after film formation, i.e. before annealing, and after annealing. In Figure 7, the 1%v-nanocomposite data for the three masses are compared to their respective intensities before annealing.

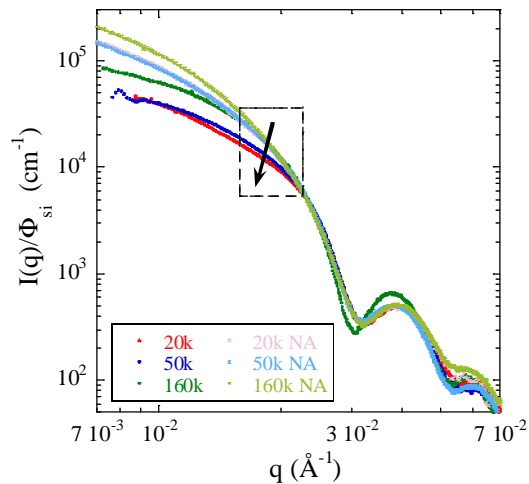

**Figure 7:** Comparison of reduced experimental scattered intensities $I(q)/\Phi_{si}$ for the 1%v-nanocomposites with and without ('NA') annealing for different matrices (20, 50 and 160 kg/mol).

The experimental curves show interesting features at intermediate q as highlighted by the dashed box in Figure 7. In this range, all samples have the same structure before annealing, corresponding to individual dispersion of silica beads (single nanoparticle form factor). It evolves into a new structure after annealing, which is the same for 20 and 50 kg/mol, as already seen in Figures 6a and 6b, and different for PEMA160. For the low masses, the structure evolves towards an aggregated state, as seen by the emergence of the correlation hole (arrow in Figure 7). On the contrary, there is no structural evolution during annealing of the PEMA160, which keeps individual dispersion. The high intensities before annealing at low q speak in favor of an initial distribution of NPs around latex beads, introducing correlations between NPs without close contact. In this range, some differences between the three masses occur. This is probably due to the fact that there is always some annealing at the end of the drying process. At high silica volume fractions, as discussed for the 10%v-nanocomposites in Figure 5b, the evolution with annealing is the same as with increasing mass, it tends towards lower mesh sizes corresponding to better dispersion. It is possible that, unlike at low concentration, the favorable silica-PEMA interaction [33, 34] explains this behavior, see e.g. [9]. In conclusion of the effect of annealing, the nanocomposite structure starts from some structure directly after drying – one may imagine a few NPs gathered in a random way around the latex beads – and evolves differently according to the matrix mass.

As a last point with structural analysis, the experimentally observed compacities $\kappa^{TEM}$ determined using eq.(5) are compared to the compacities of the model aggregates generated by the simulation. In Figure 8, this comparison is shown as a function of $N_{agg}$, where each point is an average over aggregates of same $N_{agg}$. The experimental values are given as a function of $N_{agg}^{max}$; note that choosing $N_{agg}^{min}$ would not have changed the shape of the curve, as the experimental compacities are more or less independent of aggregation number for $N_{agg}$ larger than a few units. The decrease at very low $N_{agg}$ is by continuity, as one starts from a compacity of one for an individual bead. Altogether, the experimental compacities lie on a plateau in the 8 to 12%-range. One may underline that in spite of the scattering of the data points, this is a rather precise result. Averages calculated over the functions shown in Figure 8 yield 8.4%, 11.0%, and 9.4% for chain mass 20, 50 and 160 kg/mol, respectively. This range of compacities is not compatible with those calculated using simple scaling laws ($\kappa = N_{agg}^{(Df-3/Df)}$), which assumes the same fractal dimension for aggregates of different size, see dotted line in Figure 8. This fractal scaling induces an underestimation for high $N_{agg}$, whereas for too low aggregate sizes it overestimates the compacity. In all cases, the simulation gives much more realistic values (e.g., for $N_{agg}$ = 2, the compacity is 25% by geometrical construction).

Incidentally, it is concluded that the experimentally observed fractal dimension (SAXS) is an apparent value over the aggregate distribution.

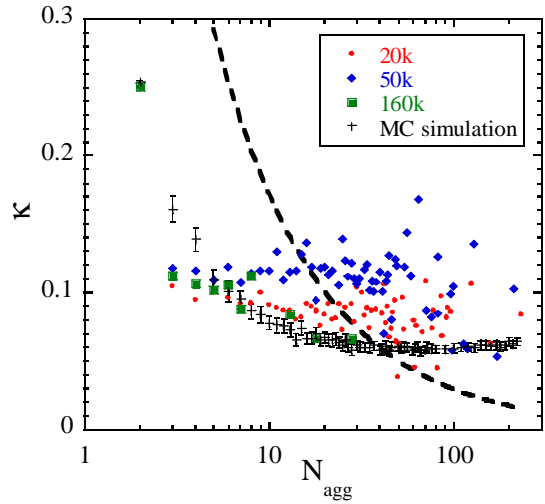

**Figure 8:** Comparison of compacity of simulated aggregates (crosses) with $\kappa^{TEM}$ deduced from TEM data following eq.(5), as a function of aggregation number $N_{agg}$ (given by $N_{agg}^{max}$ for the experimental values). Dotted line is the calculated compacity following a fractal law ($D_f = 1.7$).

For simulations, the similar trend of decreasing compacity with increasing aggregate mass is due to the same mechanism as with experiments. Above aggregation numbers of three, the numerical values are slightly lower but in the same range as the experimental ones, between 6 and 10%. This is surprisingly good as the simulated aggregate compacity is defined a priori by the aggregate construction algorithm, as outlined in the methods section. This means that real aggregates have a similar structure. It opens the way of constructing aggregates with higher local densities. One may also speculate that then the mismatch in the q-region of the correlation hole in Figure 6a and 6b could be less pronounced.

**3.2 Nanocomposite rheology**

The rheological properties of the silica-latex nanocomposites are a result of matrix properties and their modification by the presence of silica filler. It is therefore straightforward to measure the matrix first, for the three different matrix chain masses 20, 50, and 160 kg/mol. The master curves at 180°C for storage and loss moduli as a function of angular frequency are shown in Figure 9. The curves display a power law at low frequencies, ideally given by $\omega^2$ and $\omega$, for G' and G'', respectively. For G' we find an exponent of 1.5±0.2, suggesting the presence of some remaining relaxation processes in the experimental frequency window, presumably due to chain polydispersity. G'' on the other hand, is proportional to ω, the prefactor being the viscosity. In the case of the highest masses, the moduli cross at a characteristic frequency, which is also the location of a crossover to a plateau-like regime (indicated by an arrow in Figure 9). Due to polydispersity in chain mass, G' does not reach a real plateau, but increases continuously. The crossover point is shifted to higher frequencies with smaller masses, and for the lowest mass there is no more cross-over in this frequency window. The cross-over position gives the typical relaxation time, which also sets the viscosity. We have checked in Figure 9b that the viscosity of the system taken from the low-frequency prefactor of G'' is compatible with a power law $M^{3.4}$ as expected [44].

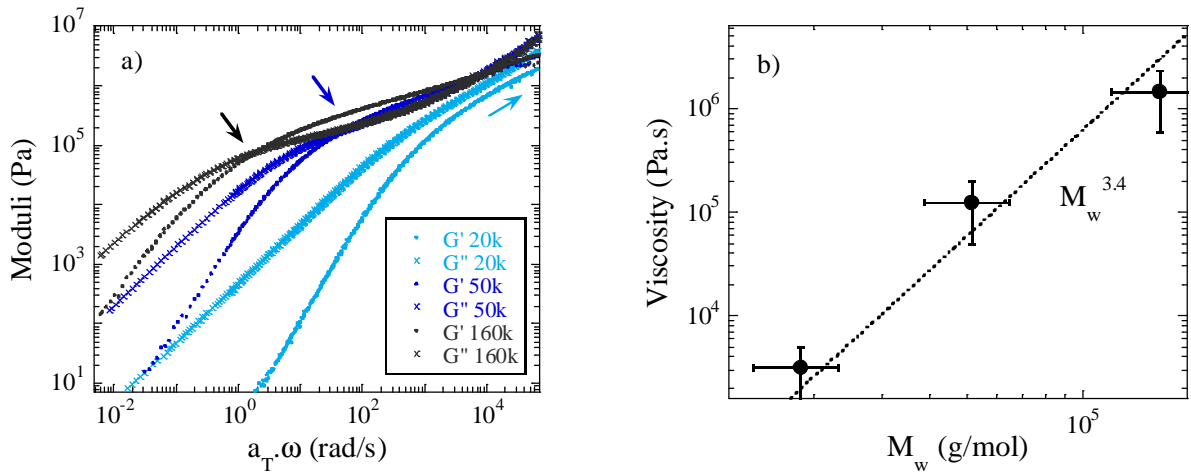

**Figure 9:** (a) Master curves for G' (dots) and G'' (crosses) as a function of angular frequency at the reference temperature of 180°C for PEMA20 (clear blue), PEMA50 (deep blue) and PEMA160 (black) matrices. (b) Viscosity extracted from the low-frequency behavior of G'' for the same matrices.

Note that the reference temperature for the construction of the master curve (180°C) corresponds to the annealing temperature of nanocomposites after film formation as described in the materials section, and we thus conclude on the impact of the chain mass on the possible maturation of the films. Indeed, the characteristic relaxation time at this temperature is increased by about three decades as the mass is increased from 20 to 160 kg/mol, causing an identical increase in viscosity (Figure 9b), and thus on the capacity of the nanocomposite structures to reorganize on the time-scale of the annealing procedure, i.e. a week. Another observation of the same change in rheology is that G'' is always greater than G' in the frequency domain of observation for the PEMA20 matrix, indicating a liquid behavior in this frequency range, whereas the other two matrices possess an elastic behavior in the plateau regime. These results thus suggest that the differences in structure as shown in Figure 1 between PEMA20, PEMA50 and PEMA160 nanocomposites have a kinetic origin. As already stated in the section on structure, our interpretation is that directly after film formation all samples start with a similar dispersion reminiscent of the colloidal organization in solution. As soon as the water is totally evaporated, the system first relaxes local concentration fluctuations, and then tends to evolve to a macroscopic phase separation of the hydrophilic silica from the hydrophobic matrix during annealing. This maturation process seems to occur in spite of the favorable silica-PEMA interactions [33, 34]. It is slowed down by the viscosity of the surrounding medium, i.e. the polymer matrix. The result is a succession of quenched silica structures, which confer different rheological properties to the final nanocomposites, as we will investigate now.

The rheological properties of nanocomposites made with the three matrices PEMA20, PEMA50, and PEMA160 and increasing amounts of silica up to 10%v have been measured over the same frequency domain as before. The loss moduli show a small increase with silica volume fraction below 10%v, and the apparition of a low-frequency plateau at 10%v parallel to G'. Therefore we concentrate here on the storage moduli G', which are presented in Figure 10. All three plots show a common dependence of G' on the silica volume fraction: at low frequencies, the curves increase with $\Phi_{si}$, whereas they tend to a common plateau (within error bars) at the highest frequencies. In agreement with other studies of acrylate reinforcement below percolation [5], the reinforcement of the high frequency modulus is quite small and mainly driven by hydrodynamic reinforcement. Striking features are found for the lowest molecular mass (Figure 10a), where the flow behavior of the pure matrix is modified progressively at intermediate frequencies as silica is added. The shape of the additional contribution suggests a process with fixed characteristic time which is caused by the silica. We will study this process in more detail below. Above 5%v of silica, the system is gelled: at 10%v there is a constant G' plateau in the low frequency part, followed by a minor increase to a second plateau at high ω. This suggests that the silica has percolated through the sample, and that the percolation threshold is located between 5 and 10%v. Such values are not uncommon for reinforced systems [10] and reflect the existence of aggregates at low chain masses.

This observation is in line with our structural analysis, which showed the formation of thick and lengthy (Figure 1) aggregates at 3%v. Such aggregates occupy a large volume as compared to their silica content, and quantified by the compacity. Indeed, as shown in Figure 8, we have found typically between 8 and 12% for $\kappa^{TEM}$ at 1%v. It is difficult to evaluate this quantity for higher volume fractions due to overlap of NPs in the TEM pictures. As the total volume fraction of aggregates is given by $\Phi_{si}/\kappa^{TEM}$, aggregates are expected to touch and interact at volume fractions below 10%v, and thus the system approaches percolation at small silica volume fractions as found by rheology.

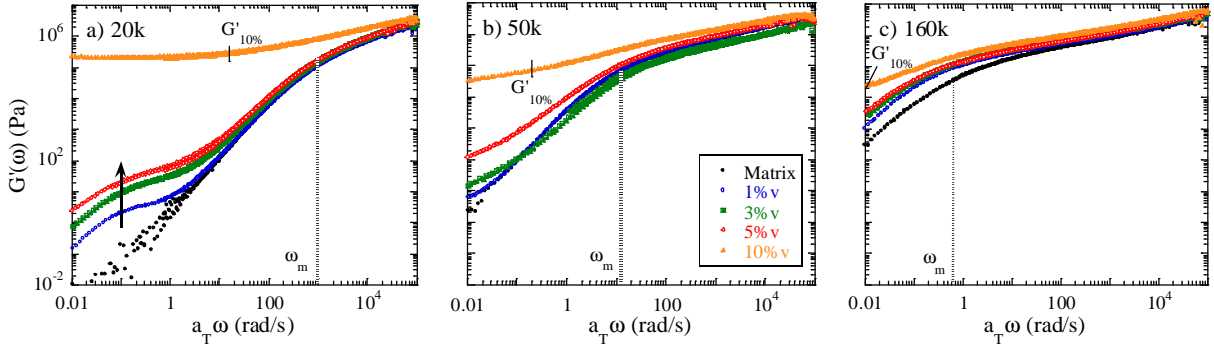

**Figure 10:** Master curves for the storage modulus G' as a function of angular frequency ω at the reference temperature of 180°C for (a) PEMA20, (b) PEMA50, and (c) PEMA160 nanocomposites. Black dots stand for matrices, blue empty dots for 1%v, green squares for 3%v, red diamonds for 5%v, and orange triangles for 10%v-composites.

The rheology of nanocomposites made with PEMA50 and PEMA160 is less rich. In Figure 10b, one can still observe the onset of the silica contribution at low frequencies as with PEMA20: the modulus increases with $\Phi_{si}$, but the process is not resolved neatly any more. The obvious explanation is that the signal of the characteristic relaxation of the matrix polymer has moved to lower frequencies, and the matrix contribution dominates the signal. For PEMA160, which had the longest relaxation time, the total modulus is even more dominated by the polymer contribution, and only a minor (but clearly visible) silica contribution can be observed at the lowest frequencies.

We now focus on the PEMA20 case shown in Figure 10a where the silica contribution can be clearly identified. It appears to be added on top of the matrix contribution, which is why we attempt to describe it as the sum of the known matrix contribution G'$_{matrix}$ and the unknown silica contribution G'$_{si}$:

$$G'(\omega) = G'_{si}(\omega, \Phi_{si}) + (1 - \Phi_{si}) G'_{matrix}(\omega) \qquad (7)$$

We can then solve eq. (7) for the first term on the r-h-s, $G'_{si}$. This process due to the silica is highlighted at low frequencies, and corresponds to a characteristic relaxation time of approximately 1/0.05 rad/s = 20 s at 180°C for PEMA20. The shape of the curves for different silica volume fractions up to 5%v is the same, and we therefore decompose $G'_{si}$ into a volume fraction independent function describing the shape $S(\omega)$, and a frequency-independent modulus $g'_{si}(\Phi_{si})$ describing the amplitude. The impact of the latter function is also illustrated by the arrow in Figure 10a.

$$G'_{si}(\omega, \Phi_{si}) = S(\omega)\, g'_{si}(\Phi_{si}) \qquad (8)$$

where $S(\omega)$ is set to one at $\omega = 0.001\, \omega_m$. The reference frequency $\omega_m$ is indicated in Figure 10. It has been set to the characteristic relaxation time of each family of nanocomposites, and depends thus on the polymer mass. The amplitude function $g'_{si}(\Phi_{si})$ is reported in Figure 11, for both PEMA20 and PEMA50. For the highest mass, it is not possible any more to extract this amplitude unambiguously. In the inset of Figure 11 the shape function $S(\omega)$ is found to superimpose nicely in the low-frequency part, suggesting that it is the same process for all low silica volume fractions.

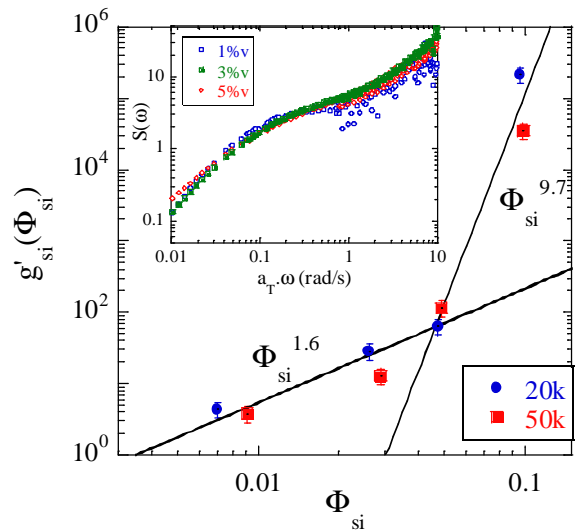

**Figure 11:** Evolution of the amplitude $g'_{si}(\Phi_{si})$ of the filler contribution for PEMA20 (circles) and PEMA50 (squares) nanocomposites. The inset shows the shape superposition of filler moduli $S(\omega)$ for the three lower volume fractions of PEMA20.

The amplitude function $g'_{si}(\Phi_{si})$ in Figure 11 follows two different regimes. Up to about 5%v, it follows a power law with exponent 1.6±0.1, followed by a much stronger increase compatible with an exponent between 9 and 10. The crossover from one to the other regime is naturally identified with the mechanical percolation of the hard silica filler in the nanocomposites. Its exact position between 5 and 10%v is unknown, however, given the strong second slope, it is probably close to 5%v. Below this threshold, aggregates are individually dispersed, and do not form a percolating hard path across the sample. The reinforcement in this regime can be understood as hydrodynamic reinforcement due to aggregates, the volume fraction of which is given by $\Phi_{si}/\kappa$. A possible scenario is 'Einsteinian' reinforcement of the form $1 + 2.5\ \Phi_{si}/\kappa$ [45]. In this view, the average aggregate compacity has to decrease to produce the total increase proportional to $\Phi_{si}^{1.6}$. As the fractal dimension is found to increase with $\Phi_{si}$ (corresponding to denser aggregates at constant size), this can only be fulfilled if the aggregate size increases in this concentration range. This is compatible with the observation of TEM pictures at 1 and 3%v (Figure 1). An alternative scenario would include higher order terms (~ $\Phi_{si}^2$), but our measurements lack the precision to distinguish between them.

After the percolation threshold, it is instructing to compare the high exponent to those predicted for colloidal gels [46] in the regime of strong links between aggregates. These models apply to space-filling assemblies of fractal aggregates, predicting a variation of the modulus proportional to $\Phi_{si}^{(3+x)/(3-D_f)}$, where x is the fractal dimension of the load-carrying aggregate backbone, and $D_f$ is the fractal dimension of the entire aggregates. The backbone being part of the fractal, one expects x to be smaller than $D_f$. Moreover, for dense fractals (i.e., $D_f$ between 2 and 3), one may postulate that x also approaches $D_f$, i.e. strongly exceeds the limiting connectivity given by Shih et al (x > 1). Using the fractal dimension estimated from SAXS ($D_f$ = 2.5±0.05, Fig.5a at 10%v), it is thus easily conceivable that x values are in the range of 2 to 2.5, making the prediction compatible with the experimentally measured exponent. In conclusion, the crowding of silica aggregates of fractal dimension observed in our structural investigation is compatible with the increase of the silica contribution to the modulus. The latter follows a power law with exponent 1.6 below the percolation threshold of ca. 5%v, and a second power law with a much greater exponent between 9 and 10, in agreement with the prediction by Shih et al [46].

# 4. DISCUSSION: IS STRUCTURE GOVERNED BY RHEOLOGICAL PROPERTIES, OR VICE VERSA?

We have seen in the preceding sections that the filler structure of the nanocomposites is governed by the mass of the polymer matrix chains. This dependency was traced back to the different viscosities of the matrices, which vary by about three orders of magnitude. The structure is thus governed by the rheology of the system during film formation and annealing. We now propose a closer look on the possibilities of spatial rearrangement of filler particles in their respective environment.

In Figure 12a, the square-root of the average mean square displacement $<r^2>$ = 6Dt of nanoparticles is plotted as a function of time t expressed as the number of days the sample is annealed. The diffusion constant D is estimated with the Stokes-Einstein equation (see e.g. [27]), and depends on the matrix viscosity given in Figure 9b. $\sqrt{<r^2>}$ thus defines a characteristic diffusion length on the timescale of the annealing procedure. It needs to be compared to the displacement typically needed to rearrange NPs, which is at least the first neighbor distance. In a perfectly dispersed nanocomposite, this distance can be estimated using a simple cubic cell model for individual NPs, and its value is given for 1%v-nanocomposites in Figure 12a. The result of this comparison can be read of the figure: for low masses, even one day of annealing (which may already take place during the initial film formation once all water is evaporated) is largely sufficient to induce reorganization. For PEMA160, however, even the full week is hardly sufficient to allow for aggregation on larger scales if low collision probabilities are taken into account. In this context, it is interesting to discuss the contribution by Jia et al [17]. They have followed aggregation as a function of polymer mass and temperature, and propose a similar analysis based on the critical diffusion time for onset of aggregation by collision (see also [25]). Differently from our experimental situation, their system possesses two competing interactions, debonding and collisions, which makes its evolution more complex.

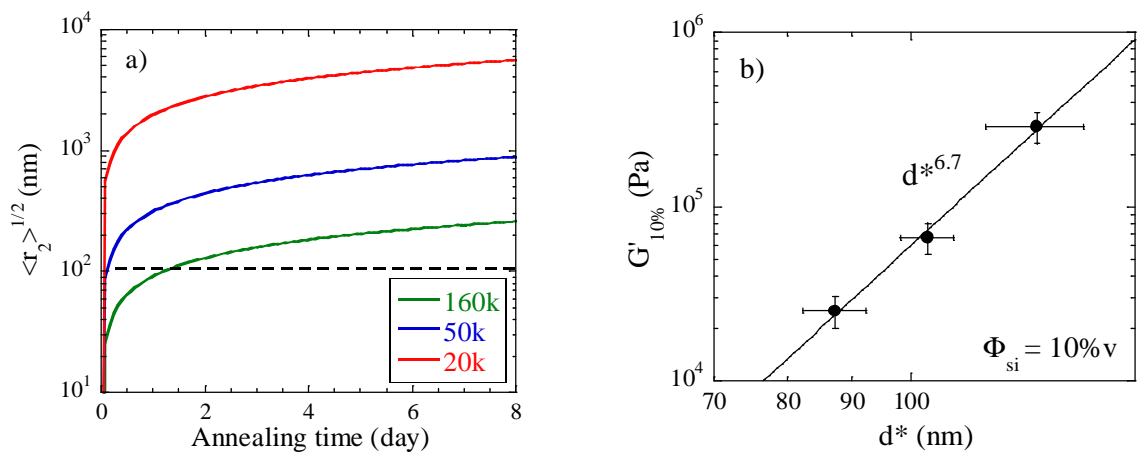

**Figure 12:** (a) Characteristic distance covered by silica particles by diffusion in the different polymer matrices during the annealing procedure as function of time. The horizontal dashed line indicates the typical distance between filler particles in perfectly dispersed 1%v-nanocomposites (cubic lattice). (b) Evolution of $G'_{10\%}$ at 0.016 $\omega_m$ as function of the silica mesh size d* found by SAXS for 10%v-nanocomposites.

To summarize, Figure 12a supports our interpretation that the matrix rheology triggers the filler structure for the resulting nanocomposite. In turn, the rheology of the final nanocomposites depends on structure. For the 10%v-nanocomposites, this is evidenced by the plot in Figure 12b of the storage modulus $G'_{10\%}$ as a function of the silica mesh size d* given in Figure 5b. For the sake of comparison between the different matrices, we have again based our evaluation on the typical relaxation frequency $\omega_m$ in Figure 10, and compared moduli at 0.016 $\omega_m$, which corresponds to the lowest frequency available for PEMA160.

The three samples contain 10%v of silica and are percolated. In spite of this similarity, the modulus is found to depend strongly on mesh size, the power law exponent being close to 7. Note that looking at the reduced modulus (i.e., divided by the corresponding matrix modulus), gives the same power-law exponent. One may add that the range in d* is rather small, the exact value of the exponent is thus subject to discussion. There is no doubt, however, on the strong dependence shown in Figure 12b. As a last comparison, it is interesting to confront it to the prediction by Shih et al [46], according to which the modulus should decrease with increasing d* due to the formation of weaker (because less dense) fractals, which is exactly opposite to our evolution. Adapting their framework to the present situation of constant volume fraction is beyond the scope of this article. An obvious contribution, different from the picture by Shih et al, comes from the thickness of the fractal branches due to volume conservation. Also, in our case there are more individual NPs in Figure 1 at high mass (small d*), which do not contribute to the branches. Thus the modulus of branches increases with d*, and so does the modulus of the percolated aggregates, as observed in Figure 12b.

## 5. CONCLUSION

We have studied the structure and rheology of silica NPs in PEMA-nanolatex matrices of different chain mass. For all volume fractions in the range from 1 to 10%v, the dispersion was found to be best for the highest chain mass, which we trace back to the slowed-down kinetics of spatial reorganization of silica in matrices of higher masses. While this is known in solvent-cast systems, we have shown here that this structure-determining parameter works also for silica-latex systems. For the lowest $\Phi_{si}$, the structure was quantified by image analysis of TEM pictures in terms of $N_{agg}$-distribution functions. With the same technique, the compacity of aggregates as a function of aggregate mass was also determined, and found to lie in the range from 8 to 12%, i.e., with a rather high degree of precision. Using Monte Carlo simulations of aggregate structure, it could be shown that the distribution functions are fully compatible with experimentally measured scattered small-angle intensities. Coupling SAXS with TEM and simulations thus validates the distribution functions of aggregation numbers. The simulations also predict similar compacities, between 6 and 10%, with a weak size dependence for $N_{agg} > 5$. Moreover, SAXS was used to follow the crowding of aggregates with silica volume fraction, up to 10%v, where network structures – which are difficult to disentangle in TEM – are found. Again, the highest mass leads to the smallest and lightest networks, which has also the lowest modulus. The rheology of the nanocomposites shows an interesting feature at low frequencies. We have identified the shape function (in the frequency domain) of the silica contribution, and analyzed the evolution of the corresponding amplitude with volume fraction. Below a percolation threshold of ca. 5%v, the modulus increases with $\sim \Phi_{si}^{1.6}$, and much stronger $\sim \Phi_{si}^{9-10}$ above it, in agreement with the prediction by Shih et al [46].

To conclude, a detailed structural and rheological analysis of a silica-latex nanocomposite system is proposed, as a function of a parameter up to now not used in such systems, the matrix chain mass. It is believed that both the results offering a new control parameter, and the coupling of techniques presented here will contribute to the understanding of the reinforcement effect in nanocomposites.

**Acknowledgements:**

We acknowledge beam-time on beamline ID2 by the European Synchrotron Radiation Facility (ESRF, Grenoble) and assistance by P. Kwasniewski. Help with TEM measurements at University of Montpellier 2 by F. Godiard and V. Richard is also acknowledged. We are thankful to J. Couve and C. Negrell-Guirao for DSC and TGA measurements, respectively. JO is indebted to the Languedoc-Roussillon region for a "Chercheur d'avenir" grant.

**Tuning structure and rheology of silica-latex nanocomposites with the molecular weight of matrix chains: a coupled SAXS-TEM-simulation approach**

Amélie Banc, Anne-Caroline Genix, Mathieu Chirat, Christelle Dupas, Sylvain Caillol, Michael Sztucki, and Julian Oberdisse

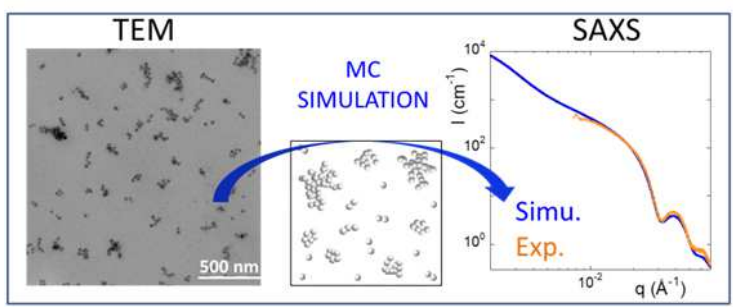